# Back to the Origins. Using Matrix Functions of Hückel Hamiltonian for Quantum Interference

Ernesto Estrada

Department of Mathematics & Statistics, University of Strathclyde, 26 Richmond Street, Glasgow, G11XH, UK

## Contents

## Prologue

When I started my first steps in theoretical chemistry at the beginning of the 1990's in Cuba I was very excited about the use of graph-theoretic ideas to represent molecules and describe their properties. While attending at a conference I had the opportunity to chat with a well-known theoretical chemist who was visiting Cuba and asking his opinion about "*chemical graph theory*". His lapidary response was: "*Chemical graph theory cannot go beyond Hückel method, and so it is dead*". Then, partially because of my ignorance and partially because of the fact that I was listening everyday many things I do not believe in, I decided to give rid of the expert advice and centre my investigation around molecules, the Hückel method and graph theory. I was in my mid 20's '*Non tamen ista meos mutabunt saecula mores: unus quisque sua noverit ire via.*' [But age shall still not change my habits: let each man be allowed to go his own way.] [1].

    A search on Google Scholar for papers published in the XXI century carried out on the 10th October 2016 has given more than 2,000 papers using the HMO method to study molecules and more than 10,000 ones using the closely related term "tight-binding Hamiltonian". It is not too bad for a method which is dead! Three remarkable examples of how alive the HMO method is in the XXI century are the following. In his 2000 Nobel lecture "The discovery of polyacetylene film: the dawning of an era

of conducting polymers" [2] Hideki Shirakawa recognized the value of HMO in determining that "*the difference between the lengths of double and single bonds decreases with increasing the conjugation and that all the bonds tend to be of equal length in an infinitely long polyene*". In 2007 Kutzelnigg [3] reviewed the HMO method for the *Journal of Computational Chemistry* with the appealing title: "*What I like about Hückel theory*" and in 2012 Ramakrishnan [4] proposed some educational tools for working with the HMO method in a paper published by the *Journal of Chemical Education*.

One of the most attractive aspects of the HMO theory resides in its connection with the mathematical theory of graphs. Graph theory [5], as it is known in mathematics, is today a very well stablished mathematical theory with connections to combinatorics, algebra, number theory, topology, theoretical computer sciences, statistical mechanics, quantum gravity, complex systems, quantum information theory, quantum chromodynamics, and of course Chemistry (for a short compilation of such relations see [6]). This connection between HMO and graph theory allows not only the numerical characterization of molecular properties, but more importantly its conceptual understanding in terms of simple molecular terms. In this sense the HMO-graph theory (HMO-GT) marriage responds positively to C. A. Coulson admonition to theoretical chemists: "*Give us insight, not numbers*" [7]. HMO-GT gives numbers and insight at the same time!

A review of the literature of the last part of the XX century shows that most of the works produced in theoretical chemistry were more on the computational side of the problem, giving more numbers and less insight to chemists. There are a few exceptions, of course. And the work of Professor Ramón Carbó-Dorca, our friend Ramón, whose work we celebrate in this book, is one of them.

## HMO and Graph Theory

In order to describe the electronic structure of conjugated molecular systems the so-called Hückel molecular orbital theory [8, 9], known in physics as the tight-binding approach [10], considers that the interaction between $N$ electrons is determined by a Hamiltonian of the following form:

$$\mathbf{H} = \sum_{n=1}^{N} \left[ -\frac{\hbar^2 \nabla_n^2}{2m} + U(r_n) + \frac{1}{2} \sum_{m \neq n} V(r_n - r_m) \right], \qquad (1)$$

where $U(r_n)$ is an external potential and $V(r_n - r_m)$ is the potential describing the interactions between electrons. Using the second quantization formalism of quantum mechanics this Hamiltonian can be written as:

$$\hat{H} = -\sum_{ij} t_{ij} \hat{c}_i^\dagger \hat{c}_j + \frac{1}{2} \sum_{ijkl} V_{ijkl} \hat{c}_i^\dagger \hat{c}_k^\dagger \hat{c}_l \hat{c}_j, \qquad (2)$$

where $\hat{c}_i^\dagger$ and $\hat{c}_i$ are 'ladder operators', $t_{ij}$ and $V_{ijkl}$ are integrals which control the hopping of an electron from one site to another and the interaction between electrons, respectively. They are usually calculated directly from finite basis sets [11].

In the HMO approach for studying solids and conjugated molecules, the interaction between electrons is neglected and $V_{ijkl} = 0, \forall i, j, k, l$. This method can be seen as very drastic in its approximation, but let us think of the physical picture behind it [2, 12]. We concentrate our discussion on alternant conjugated molecules in which single and double bonds alternate. Consider a molecule like benzene in which every carbon atom has an $sp_2$ hybridization. The frontal overlapping $sp_2-sp_2$ of adjacent carbon atoms create very stable $\sigma$-bonds, while the lateral overlapping $p-p$ between adjacent carbon atoms creates very labile $\pi$-bonds. Thus it is clear from the reactivity of this molecule that a $\sigma-\pi$ separation is plausible and we can consider that our basis set consists of orbitals centred on the particular carbon atoms in such a way that there is only one orbital per spin state at each site. Then we can write the Hamiltonian of the system as:

$$\hat{H}_{tb} = -\sum_{ij} t_{ij} \hat{c}^\dagger_{i\rho} \hat{c}_{i\rho}, \tag{3}$$

where $\hat{c}^{(\dagger)}_{i\rho}$ creates (annihilates) an electron with spin $\rho$ in a $\pi$ (or other) orbital centred at the atom $i$. We can now separate the in-site energy $\alpha_i$ from the transfer energy $\beta_{ij}$ and write the Hamiltonian as

$$\hat{H}_{tb} = \sum_{ij} \alpha_i \hat{c}^\dagger_{i\rho} \hat{c}_{i\rho} + \sum_{\langle ij \rangle \rho} \beta_{ij} \hat{c}^\dagger_{i\rho} \hat{c}_{i\rho}, \tag{4}$$

where the second sum is carried out over all pairs of nearest-neighbours. Consequently, in a conjugated molecule or solid with $N$ atoms the Hamiltonian (2.3) is reduced to an $N \times N$ matrix,

$$H_{ij} = \begin{cases} \alpha_i & \text{if } i = j \\ \beta_{ij} & \text{if } i \text{ is connected to } j \\ 0 & \text{otherwise.} \end{cases} \tag{5}$$

Due to the homogeneous geometrical and electronic configuration of many systems analyzed by this method we may take $\alpha_i = \tilde{\alpha}, \forall i$ (Fermi energy) and $\beta_{ij} = \tilde{\beta} \approx -2.70 eV$ for all pairs of connected atoms. Thus,

$$\hat{H} = \tilde{\alpha} I + \tilde{\beta} A \tag{6}$$

where $I$ is the identity matrix, and $A$ is a matrix whose entries are defined as

$$A_{ij} = \begin{cases} 1 & \text{if atom } i \text{ is bonded to atom } j \\ 0 & \text{otherwise.} \end{cases} \tag{7}$$

Let us now introduce the following concepts. Let us consider a finite set $V = \{v_1, v_2, \cdots, v_N\}$ of unspecified elements and let $V \otimes V$ be the set of all ordered pairs $[v_i, v_j]$ of the elements of $V$. A relation on the set $V$ is any subset $E \subseteq V \otimes V$. The relation $E$ is symmetric if $[v_i, v_j] \in E$ implies $[v_j, v_i] \in E$ and it is reflexive if $\forall v \in V, [v, v] \in E$. The relation $E$ is antireflexive if $[v_i, v_j] \in E$ implies $v_i \neq v_j$. Now let us define a *simple graph* as the pair $G = (V, E)$, where $V$ is a finite set of nodes (also known as vertices), and $E$ is a symmetric and antireflexive relation on $V$, whose elements are known as the edges or links of the graph [13].

Thus, if in any conjugated molecule we consider every non-hydrogen atom as a vertex, and every covalent bond, excluding those involving hydrogen, as an edge, we have a direct map between a molecule and a graph. These graphs are known as hydrogen-deleted graphs, or simply as molecular graphs. In this context, the matrix $A$ is the adjacency matrix of the graph, which represents the adjacency relation between the nodes of the corresponding graph. Then, the adjacency matrix of the molecular graph and the HMO Hamiltonian have the same eigenfunctions $\varphi_j$ and their eigenvalues are simply related by:

$$\hat{H}\varphi_i = E_i A, \ A\varphi_i = \lambda_i \hat{H}, \ E_i = \tilde{\alpha} + \tilde{\beta}\lambda_i \tag{8}$$

Hence everything we have to do in the analysis of the electronic structure of molecules or solids that can be represented by an HMO Hamiltonian, is to study the spectra of the graphs associated with them. The study of spectral properties of graphs represents an entire area of research in algebraic graph theory. The spectrum of a matrix is the set of eigenvalues of the matrix together with their multiplicities. For the case of the adjacency matrix let $\lambda_1(A) \geq \lambda_2(A) \geq \cdots \geq \lambda_N(A)$ be the distinct eigenvalues of $A$ and let $m(\lambda_1(A)), m(\lambda_2(A)), \cdots, m(\lambda_N(A))$ be their algebraic multiplicities, i.e., the number of times each of them appears as an eigenvalue of $A$. Then the spectrum of $A$ can be written as [13]

$$SpA = \begin{pmatrix} \lambda_1(A) & \lambda_2(A) & \cdots & \lambda_N(A) \\ m(\lambda_1(A)) & m(\lambda_2(A)) & \cdots & m(\lambda_N(A)) \end{pmatrix}. \tag{9}$$

The total $\pi$ (molecular) energy is given by

$$E = \tilde{\alpha}n_e + \tilde{\beta}\sum_{i=1}^{N} g_i \lambda_i, \tag{10}$$

where $n_e$ is the number of $\pi$-electrons in the molecule and $g_j$ is the occupation number of the $j$-th molecular orbital. For neutral conjugated systems in their ground state we have [14]

$$E = \begin{cases} 2\sum_{j=1}^{n/2} \lambda_j & n \text{ even,} \\ 2\sum_{j=1}^{(n+1)/2} \lambda_j + \lambda_{(j+1)/2} & n \text{ odd.} \end{cases} \tag{11}$$

Because an alternant conjugated hydrocarbon has a bipartite molecular graph, i.e., a graph whose set of vertices can be split into two disjoint sets $V_1$ and $V_2$, $V = V_1 \bigcup V_2$ : $\lambda_i = -\lambda_{N-i+1}$ for all $j = 1, 2, \ldots, N$. In a few molecular systems the spectrum of the adjacency matrix is known. For instance [2], we have

i) Polyenes $C_N H_{N+2}$

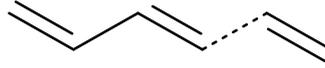

$$\lambda_j(A) = 2\cos\left(\frac{\pi j}{N+1}\right), \ j = 1, \ldots, N, \tag{12}$$

    ii)    Cyclic polyenes $C_N H_N$

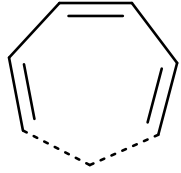

$$\lambda_j(A) = 2\cos\left(\frac{2\pi j}{N}\right), \ j = 1, \ldots, N, \ \lambda_i = \lambda_{N-i} \tag{13}$$

    iii)    Linear polyacenes, $N = 4Z + 2$

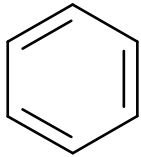 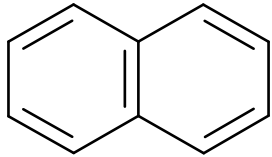 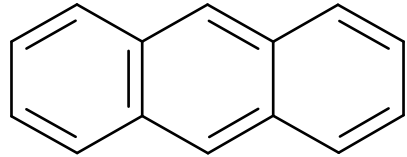

            $Z = 1$                $Z = 2$                $Z = 3$

$$\lambda_r(A) = 1; \lambda_s(A) = -1;$$
$$\lambda_k(A) = \pm\frac{1}{2}\left\{1 \pm \sqrt{9 \pm 8\cos\frac{k\pi}{Z+1}}\right\}, k = 1, \ldots, Z, \tag{14}$$

where all four combinations of signs have to be considered

## Matrix Functions of the HMO Hamiltonian

    There are several alternative definitions of a matrix function. If the matrix $\hat{H}$ is diagonalizable we can express it as $\hat{H} = V\Lambda V^{-1}$, where $V$ is a matrix whose columns are the (orthonormal) eigenvectors of $\hat{H}$ and $\Lambda$ is a diagonal matrix of eigenvalues. Then, a function of $\hat{H}$ is defined by [15]

$$f(\hat{H}) = V\begin{bmatrix} f(\lambda_1) & \cdots & 0 \\ \vdots & \ddots & \vdots \\ 0 & \cdots & f(\lambda_n) \end{bmatrix} V^{-1}. \tag{15}$$

The function $f(\hat{H})$ has the following Taylor expansion:

$$f(\hat{H}) = f(0) + f'(0)\hat{H} + f''(0)\frac{\hat{H}^2}{2!} + \cdots, \tag{16}$$

and can be represented by using the Cauchy integral formula [15]:

$$f(\hat{H}) = \frac{1}{2\pi i} \oint_C f(z)(zI - \hat{H})^{-1} dz, \tag{17}$$

where $f$ is analytic on and inside the closed contour that encloses the eigenvalues of $A$.

As we have seen in the previous section there is an identity correspondence between the HMO Hamiltonian and the adjacency matrix of the molecular graph: $\hat{H} = -A$. Then, let us consider that the molecule is submerged into a thermal bath of inverse temperature $\beta = (k_B T)^{-1}$. We are interested in knowing what is the probability that the molecule is found in an energy state $E_j = -\lambda_j$. Assuming, as usual, a Boltzmann distribution we get

$$p_j = \exp(-\beta E_j) / \sum_i \exp(-\beta E_i). \tag{18}$$

The denominator of this expression is known as the electronic partition function of the molecule [16, 17]

$$Z = \sum_i \exp(-\beta E_i). \tag{19}$$

Then, using our equivalence with the adjacency matrix we have

$$p_j = \exp(\beta \lambda_j) / Z. \tag{20}$$

It is easy to see that the partition function is

$$Z = Tr \exp(-\beta \hat{H}) = Tr \exp(\beta A), \tag{21}$$

where $Tr$ represents the trace of the matrix, and

$$\exp(\beta A) = I + \beta A + \frac{(\beta A)^2}{2!} + \cdots + \frac{(\beta A)^k}{k!} + \cdots, \tag{22}$$

which according to our previous definition is a matrix function of the adjacency matrix of the molecular graph [18, 19].

In chemical graph theory this partition function is typically represented as $Z = EE(G, \beta)$ and it is known as the Estrada index of the graph, in particular for $\beta = 1$ (for recent reviews see [20, 21]). It is well-known that

$$\exp(\beta A) = \sinh(\beta A) + \cosh(\beta A), \tag{23}$$

where

$$\sinh(\beta A) = \sum_{k=0}^{\infty} \frac{(\beta A)^{2k+1}}{(2k+1)!}, \tag{24}$$

$$\cosh(\beta A) = \sum_{k=0}^{\infty} \frac{(\beta A)^{2k}}{(2k)!}, \tag{25}$$

are two other matrix functions. In bipartite graphs, such as those representing conjugated molecules, $\sinh(\beta A) = 0$ due to the lack of odd cycles in their structures. Thus,

$$Z = EE(G,\beta) = Tr\cosh(\beta A). \tag{26}$$

There are several other matrix functions, such as trigonometric functions, sign function, $\psi$-functions, and others. The interested reader can consult the following references for more details [22].

## Quantum Interference

The phenomenon of quantum interference (QI) consists of a significant reduction of the conductance through the bonds of a molecule to which electrodes have been connected to specific atoms [23, 24]. In conjugated polycyclic compounds, QI controls site-dependent electron transport as corroborated experimentally by different techniques [25-30]. QI has attracted the attention of theorists who have developed different kind of approaches to explain this phenomenon. They include local atom to atom transmission [31], the use of simple Hückel molecular orbital (HMO) theory [32], spectral methods [33], graph-theoretic concepts [34] and graphical approaches [35-37]. An example of the last approach is the method developed by Markussen et al. [35], which provides a direct link between QI and the topology of various alternant π systems including *meta*-linked benzene derivatives, anthraquinone derivatives, and cross-conjugated molecules. Such rules can be resumed as follows: (i) Two sites can be connected by a path if they are nearest neighbors. (ii) At all internal sites, i.e., sites other than 1 and *N*, there is one incoming and one outgoing path. It is now straightforward to show that the condition for complete destructive interference is fulfilled if it is *not* possible to connect the external sites 1 and *N* by a continuous chain of paths and at the same time fulfil the rules (i) and (ii). On the other hand, if such a continuous path can be drawn, then the condition $\det_{1N}(H_{mol}) = 0$ is not fulfilled and a transmission antiresonance does not occur at the Fermi energy.

Recent results obtained by Xia et al. [38] have challenged some of the theories proposed for the QI effects in conjugated polycyclic compounds. In particular they obtained azulene derivatives containing gold-binding groups at different points of connectivity within the azulene core. By comparing paths through the 5-membered ring, 7-membered ring, and across the long axis of azulene they have found that simple models, such as the one developed by Markussen et al. [35], cannot be used to predict quantum interference characteristics of nonalternant hydrocarbons. In particular, azulene derivatives that are predicted to exhibit destructive interference based on widely accepted atom-counting models show a significant conductance at low biases. In Figure 1 we illustrated the predictions made by using Markussen et al. [35] graphical model for 1,3- and 5,7-azulene, which are predicted to display QI but are observed to transmit current through the electrodes. For the sake of comparison the application of the same graphical rules are displayed for naphthalene 1,3- and 2,7-naphthalene.

Naphthalene                                    Azulene

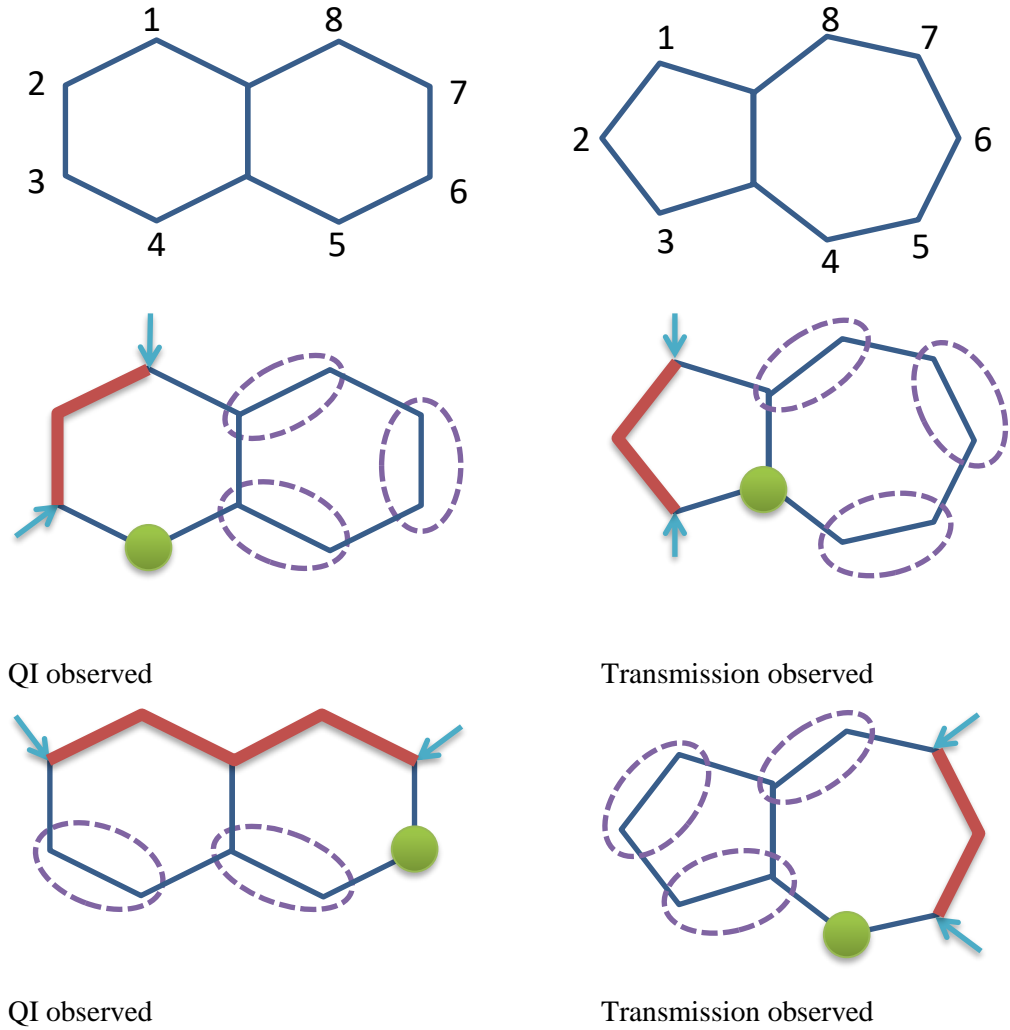

**Figure 1**. Illustration of Markussen et al. [35] rules for naphthalene and azulene. The rules predict that there is QI in 1,3- and 5,7-azulene, but transmission has been experimentally observed for these azulene derivatives.

My interest in this topic has been motivated by a meeting in Glasgow with Professor Roald Hoffmann (Nobel Prize winner in Chemistry, 1982) who introduced me to the topic. Hoffmann et al. [39-42] have used the inverse of the HMO Hamiltonian to account for QI in polycyclic conjugated hydrocarbons. The idea is to consider the Green's function

$$G_{rs}^{(0)}(E_F) = \sum_k \frac{C_{rk}C_{sk}^*}{E_F - \varepsilon_k + i\eta}, \qquad (27)$$

for which, without any loss of generality, we will consider $\eta = 0$. The Green function can be written in matrix form as $G^0(E_F) = (E_F I - \hat{H})^{-1}$, where $I$ is the identity matrix. Then, by considering the case $G^0(E_F = 0) = -\hat{H}^{-1}$, QI is predicted for the pair of atoms for which $G_{rs}^{(0)}(E_F = 0) = 0$. This model displays many beautiful and interesting properties, such as mathematical elegance, simplicity and good predictability in alternant conjugated hydrocarbons. However, when applied to azulene derivatives, $G^0\left(E_F = 0\right)$ predicts QI for the 5,7-azulene, which has been observed to transmit the electric current.

Here we propose a simple model based on HMO theory that not only predict correctly QI in conjugated (alternant and nonalternant) molecules, but also allows the development of simple graphical rules for predicting QI in these molecules.

## An HMO Matrix-Function Model for QI

We start by considering the Green function (1) by using the HMO Hamiltonian. Here we consider the following modification of the Hamiltonian

$$\hat{H}^T = \hat{H} - \gamma^2 P, \tag{28}$$

where $P$ represents a small perturbation of the Hückel Hamiltonian based on electron hopping beyond the first nearest-neighbour in the molecule and $\gamma$ is a parameter that controls the strength of the perturbation (notice that $-\gamma^2$ is used to guarantee that this contribution is always negative). It is well-known that the nondiagonal entries of $\hat{H}^k$ count the number of walks (hops) of length $k$ between the corresponding pair of atoms. A walk is a sequence of (not necessarily different) atoms and bonds starting at one node and ending at another one. For instance, $\left(\hat{H}^3\right)_{rs}$ gives the number of walks of length 3 between the atoms $r$ and $s$. Usually such perturbation is accounted for by using a McLaurin series expansion of the Hamiltonian: $P = \kappa^2 \hat{H}^2 + \kappa^3 \hat{H}^3 + \cdots$. Here we introduce the following modification to the perturbation theory of the Hamiltonian. We consider that $P$ has a closed form in order to avoid approximation problems with the power series, such that $\hat{H}^T = \hat{H} - \gamma^2 f\left(\hat{H}\right)$, where $f\left(\hat{H}\right)$ is a matrix function of the Hamiltonian. The function $f\left(\hat{H}\right)$ has to fulfil a few requirements. First, we need to give more weight to the energy levels which are close to the frontier orbitals, which are the ones more involved in the electron conduction (see further). Second, if $\lambda$ is an eigenvalue of $\hat{H}$ we require that $f\left(-\lambda\right) = -f\left(\lambda\right)$. This condition is imposed by the observed fact that when the eigenvalues of $\hat{H}$ are symmetric around the Fermi level the transmission spectra is symmetric around zero. These two conditions are fulfilled by the inverse of the Hamiltonian:

$$P = \frac{(-1)^{N-1}}{\det\left(\hat{H}\right)}\left(\hat{H}^{N-1} + c_{N-2}\hat{H}^{N-2} + \cdots + c_1 I\right) = \hat{H}^{-1}, \tag{29}$$

where $\det\left(\hat{H}\right)$ is the determinant of $\hat{H}$, which is assumed to be nonzero, $N$ is the number of carbon atoms and $I$ is the identity matrix.

Now we can write the Green's function of $\hat{H}^T = \hat{H} - \gamma^2 \hat{H}^{-1}$ as

$$G^T(E_F) = (E_F I - \hat{H}^T)^{-1} = (E_F I - \hat{H} + \gamma^2 \hat{H}^{-1})^{-1}, \qquad (30)$$

which for $E_F = 0$ reduces to $\mathscr{F} = G^T(E_F = 0) = (-\hat{H} + \gamma \hat{H}^{-1})^{-1}$. Using the property of the inverse of two non-singular matrices: $(AB)^{-1} = B^{-1} A^{-1}$, we can write $\mathscr{F}$ as

$$\mathscr{F} = \hat{H}(\gamma^2 I - \hat{H}^2)^{-1}. \qquad (31)$$

The term in the parenthesis represents a renormalization of the Hamiltonian that folds the original spectrum around the Fermi energy. That is, a Hamiltonian renormalization that places the frontier orbitals (and those close to them) at the bottom of the energy scale, while the orbitals with the highest and lowest energy are placed at the top of this scale. This renormalization is obtained by considering the squared Hamiltonian, such that $\hat{H}^2 \psi = \varepsilon^2 \psi$. The '*squared Hamiltonian trick*' has been used previously for studying different molecular systems ranging from quasicrystals to doped graphene [43-45]. This renormalization agrees with the idea that the highest contribution to the Green's function is usually made by the frontier orbitals, i.e., those at the middle of the spectrum and thus close to $E_F$. This, of course, includes the highest occupied molecular orbital (HOMO) and the lowest unoccupied molecular orbital (LUMO). In this case, as analyzed by Yoshizawa et al. [46], the sign of the product $C_{rHOMO} C^*_{sHOMO}$ and of $C_{rLUMO} C^*_{sLUMO}$ mainly determine whether there is transmission between the atoms $r$ and $s$ or there is QI. For instance, if $\text{sgn}(C_{rHOMO} C^*_{sHOMO}) \neq \text{sgn}(C_{rLUMO} C^*_{sLUMO})$ the contributions of the frontier orbitals are enhanced and there is transmission. On the other hand, if $\text{sgn}(C_{rHOMO} C^*_{sHOMO}) = \text{sgn}(C_{rLUMO} C^*_{sLUMO})$ the contributions from these orbitals are cancelled out which may produce QI among this pair (we recall that the contribution of other orbitals is being neglected in this analysis).

For any pair of atoms in a molecule we obtain

$$\mathscr{F}_{rs} = \sum_k \frac{\varepsilon_k C_{rk} C^*_{sk}}{\gamma^2 - \varepsilon_k^2}. \qquad (32)$$

Obviously $\mathscr{F}$ is just the original Green's function of the Hückel model for the limit $\gamma^2 \to 0$ (no perturbation of the Hamiltonian):

$$\lim_{\gamma \to 0} \mathscr{F} = \lim_{\gamma \to 0} \hat{H}(\gamma^2 I - \hat{H}^2)^{-1} = -\hat{H}^{-1} = G^0(E_F = 0). \qquad (33)$$

We notice that the function $\mathscr{F} = \hat{H}\left(\gamma^2 I - \hat{H}^2\right)^{-1}$, $0 < \gamma^2 < \min_j \varepsilon_j^2$, always exists. Thus, the condition imposed before that the Hamiltonian needs to be non-singular is now removed. In fact, it exists for any $\gamma^2 \notin sp\{\hat{H}\}$, where $sp\{\hat{H}\}$ is the set of eigenvalues of $\hat{H}$. However, for $\gamma^2 > \max_j \varepsilon_j^2$ the function $\mathscr{F}$ is positive, which means that the sign of $\mathscr{F}_{rs}$ is always positive, destroying one of the nice characteristics of $-\hat{H}^{-1}$ as reported by Tsuji et al. [41]. Thus, we will consider here the case when $0 < \gamma^2 < \min_j \varepsilon_j^2$. In this case, the nondiagonal entries of $\mathscr{F}$ indicates whether there is QI or transmission according to the following:

i. If $\langle r|\mathscr{F}|s\rangle = 0$ there is QI between $r$ and $s$.

ii. If $\langle r|\mathscr{F}|s\rangle \neq 0$ there is current transmission between $r$ and $s$.

As a first example of the potentialities of the current method we study here the transmission through the atoms 2 and 3 of 1,3-butadiene. This molecule was studied by Solomon et al. [47], where it was found that when using Hückel method QI phenomenon is observed between these two atoms. However, when more sophisticated MDE many-body theory is used there is a splitting of the central super-node, a certain level of transmission is observed for this connection (see Fig. 9 in Solomon et al. [47] paper). The same effect is observed here by considering $\left(\mathscr{F}_{2,3}\right)^2$ (normalized to one) as the transmission for different values of energy (see Fig. 2). The structural interpretation of this result is given in the next section.

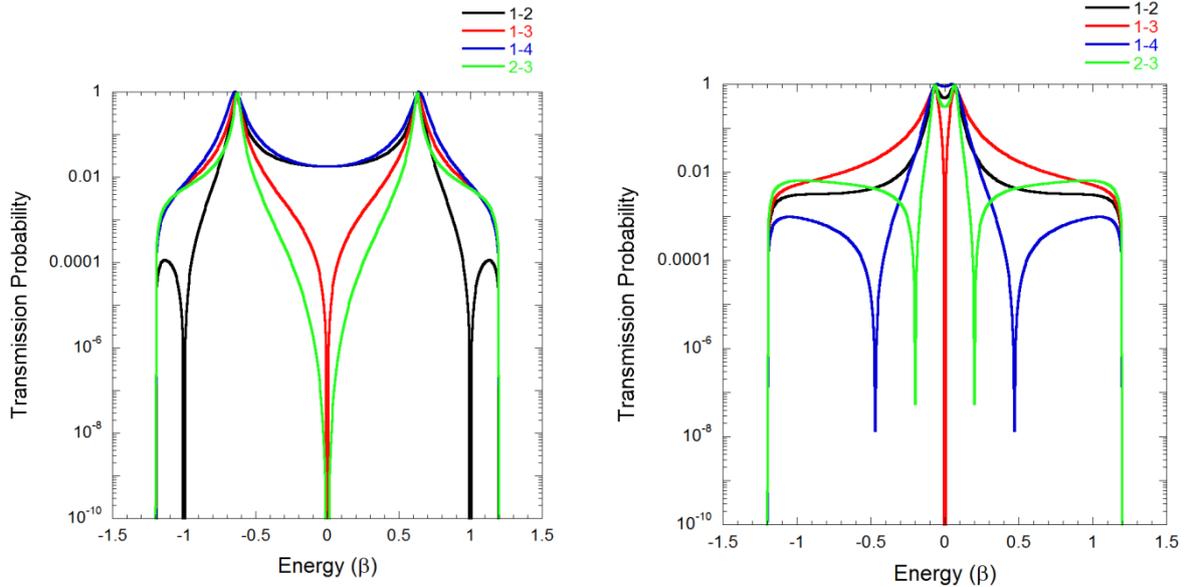

**Figure 2**. Illustration of the transmission between the different pairs of atoms in 1,3-butadiene using the Hückel method $-\hat{H}^{-1}$ (left) and $\left(\mathscr{F}_{2,3}\right)^2$ (right) using $\gamma^2 = 0.425$. The plots are in linear-log scale.

# Rules for QI

We start by expanding the function $\mathcal{F}$ as a power series of the Hamiltonian. We derive here the power series expansion of the function $\mathcal{F}$. Let us use the Cayley-Hamilton for the inverse matrix function $\left(\gamma^2 I - \hat{H}^2\right)^{-1}$:

$$\left(\gamma^2 I - \hat{H}^2\right)^{-1} = \frac{(-1)^{N-1}}{\det\left(\gamma^2 I - \hat{H}^2\right)}\left[\left(\gamma^2 I - \hat{H}^2\right)^{N-1} + c_{N-2}\left(\gamma^2 I - \hat{H}^2\right)^{N-2} + \cdots + c_1 I\right]. \tag{34}$$

First, let us consider the term multiplying the whole parenthesis in (S1). If the number of carbon atoms is even then $(-1)^{N-1} = -1$. For $0 < \gamma^2 < \min_j \varepsilon_j^2$ all eigenvalues of $\gamma^2 I - \hat{H}^2$ are negative, i.e., it is negative definite. Then, $\det\left(\gamma^2 I - \hat{H}^2\right) > 0$. Thus, the term $(-1)^{N-1}/\det\left(\gamma^2 I - \hat{H}^2\right) < 0$. Let us replace this term by $-Q$, where $Q \in \mathfrak{R}^+$.

Now, we will use the binomial theorem for the power terms of (S1):

$$(x+y)^n = \sum_{k=0}^{n}\binom{n}{k}x^k y^{n-k}. \tag{35}$$

For instance, for benzene these terms become

$$\left(\gamma^2 I - \hat{H}^2\right)^2 = \hat{H}^4 - 2\gamma^2 \hat{H}^2 + \gamma^4, \tag{36}$$

$$\left(\gamma^2 I - \hat{H}^2\right)^3 = -\hat{H}^6 + 3\gamma^2 \hat{H}^4 - 3\gamma^4 \hat{H}^2 + \gamma^6, \tag{37}$$

$$\left(\gamma^2 I - \hat{H}^2\right)^4 = \hat{H}^8 - 4\gamma^2 \hat{H}^6 + 6\gamma^4 \hat{H}^4 - 4\gamma^6 \hat{H}^2 + \gamma^8, \tag{38}$$

$$\left(\gamma^2 I - \hat{H}^2\right)^5 = -\hat{H}^{10} + 5\gamma^2 \hat{H}^8 - 10\gamma^4 \hat{H}^6 + 10\gamma^6 \hat{H}^4 - 5\gamma^8 \hat{H}^2 + \gamma^{10}, \tag{39}$$

$$\left(\gamma^2 I - \hat{H}^2\right)^6 = \hat{H}^{12} - 6\gamma^2 \hat{H}^{10} + 5\gamma^4 \hat{H}^8 - 20\gamma^6 \hat{H}^6 + 15\gamma^8 \hat{H}^4 - 6\gamma^{10} \hat{H}^2 + \gamma^{12}. \tag{40}$$

Grouping together all the similar terms in the binomial expansion of the terms in (34) we get

$$\left(\gamma^2 I - \hat{H}^2\right)^{-1} = Q\left[\hat{H}^{2(N-1)} - \vartheta_{N-2}c_{N-2}\hat{H}^{2(N-2)} - \cdots + \vartheta_2 c_2 \hat{H}^2 - \alpha_1 c_1 I\right]. \tag{41}$$

According to the Cayley-Hamilton theorem the coefficients $c_k$ are the ones of the characteristic polynomial of $\gamma^2 I - \hat{H}^2$. We use here a result by Brooks [48] which proves that the characteristic polynomial of a given matrix can be expressed as (42)

$$P(x) = x^N - \left[\sum_{\text{sets of 1}}\lambda\right]x^{N-1} + \left[\sum_{\text{sets of 2}}\lambda\lambda\right]x^{N-2} - \cdots + (-1)^k\left[\sum_{\text{sets of k}}\lambda\lambda\cdots\lambda\right]x^{N-k} + \cdots (-1)^N\left[\sum_{\text{sets of n}}\lambda\lambda\cdots\lambda\right].$$

where $\lambda$ is an eigenvalue of the corresponding matrix. Then, (43)

$$\left(\gamma^2 I - \hat{H}^2\right)^{-1} = Q\left[\hat{H}^{2(N-1)} + \vartheta_{N-2}\left[\sum_{\text{sets of } 1}\lambda\right]\hat{H}^{2(N-2)} - \cdots + \vartheta_2\left[\sum_{\text{sets of } n-2}\lambda\lambda\cdots\lambda\right]\hat{H}^2 - \vartheta_1\left[\sum_{\text{sets of } n-1}\lambda\lambda\cdots\lambda\right]I\right]$$

Because $\gamma^2 I - \hat{H}^2$ is negative definite, i.e., all $\lambda < 0$ we have that

$$\left[\sum_{\text{sets of } k}\lambda\cdots\lambda\right]\begin{cases}< 0 \text{ if } k \text{ is odd} \\ > 0 \text{ if } k \text{ is even}\end{cases}. \tag{44}$$

Thus,

$$\left(\gamma^2 I - \hat{H}^2\right)^{-1} = Q\left[\hat{H}^{2(N-1)} - \delta_{N-2}\hat{H}^{2(N-2)} + \cdots - \delta_2\hat{H}^2 + \delta_1 I\right], \tag{45}$$

where $\delta_k$ groups all the coefficients. Then, we finally obtain the power series for $\mathcal{F}$:

$$\mathcal{F} = \hat{H}\left(\gamma^2 I - \hat{H}^2\right)^{-1} = Q\left[\hat{H}^{2(N-1)} - \delta_{N-2}\hat{H}^{2(N-2)} + \cdots - \delta_2\hat{H}^2 + \delta_1 I\right]. \tag{46}$$

where $N$ is the number of carbon atoms, Q is a positive number and the coefficients $\delta_k$ depends on $\gamma^2$ and on the coefficients of the Cayley-Hamilton expansion.

Clearly, there are only odd-length walks contributing to $\mathcal{F}$. This means that $\mathcal{F}_{rs} = 0$ if and only if there are no walks of odd length between the atoms $r$ and $s$. Obviously, the existence of a walk of odd length implies the existence of an odd-length path connecting the two nodes, i.e., a walk in which all atoms and bonds are distinct. Consequently, the rules of QI in conjugated molecules can be formulated as:

i. There is QI between the atoms $r$ and $s$ if and only if there is no path of odd length connecting them.

ii. There is transmission between the atoms $r$ and $s$ if and only if there is at least one path of odd length connecting them.

In order to make a clearer interpretation of these results we start by considering an electric circuit formed by the source (electrodes) connected to a pair of atoms and the alternant bond forming resistors connected in series between the two electrodes. As usual in electrical circuit the current flows from the negative pole of the source to the positive one through the wires and resistors. The last ones are polarized in the direction of the current. We consider here that a bond represents a resistor in which the $\pi$-electrons at the corresponding atoms have opposite spin. In addition, according to the spin alternation rule, which states that the singlet spin pairing is preferred solely between sites in different subsets, the free valences on the starred and unstarred sites might be identified with "*up*" and "*down*" spin. Let us assume that the polarity of the resistor in the direction of the current implies that the spin of the two electrons in the bond are respectively *down-up* ($\downarrow - \uparrow$). Then, there is a current flow between the two electrodes if and only if there is at least one alternant sequence of spins of the form: $\downarrow - \uparrow \cdots \downarrow - \uparrow$. This is only possible if the path connecting the two electrodes is of even length. We assume that if there is at least one of such paths the current will use it to flow from one electrode to the other. Obviously, there is QI when no such path exists between the two electrodes. Notice that all paths should be taken into account (we will illustrate this later). In Figure 2 we illustrate the two situations described before.

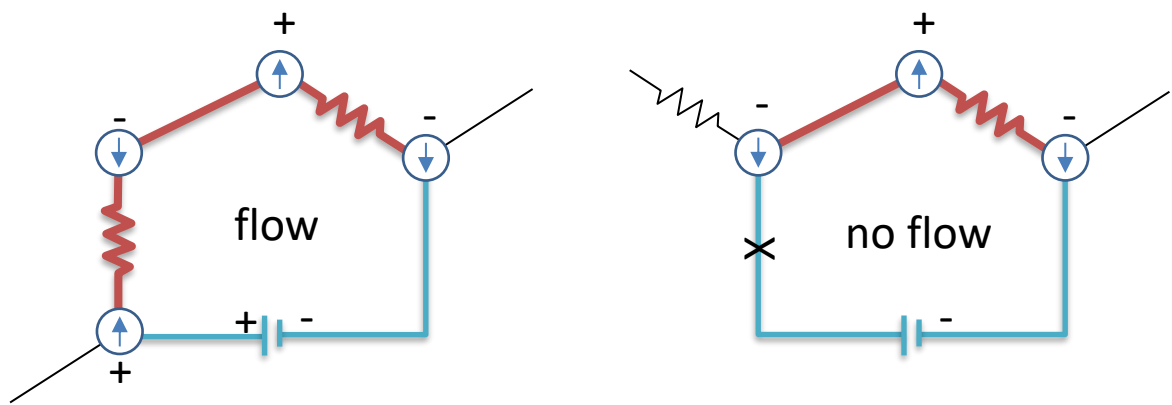

**Figure 2**. Schematic representation of the structural rules for the existence of current transmission or QI between a pair of atoms. Flow exists if and only if the polarity of the circuit is not altered as illustrated on the left-hand side of the picture. QI appears when the polarity at the starting and ending points are the same, which impedes the flow of current in the circuit.

In an odd cycle there is at least one path of odd length between any pair of atoms. Thus, $\langle r|\mathcal{F}|s\rangle \neq 0$ for every $r$ and $s$ and there is no QI in such molecules. The typical example is azulene (see Figure 3). Although it is an alternant conjugated molecule, it is formed by two odd-length cycles and there is always an odd length path between every pair of atoms and consequently a current can be transmitted between them.

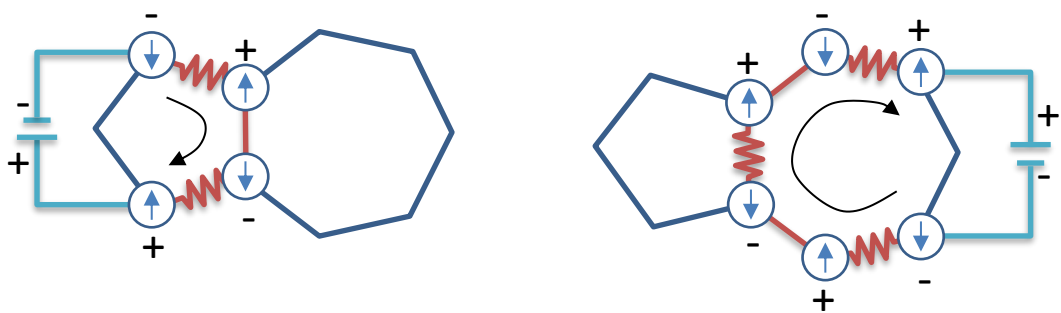

**Figure 3**. Illustration of the prediction of current transmission in 1,3- and 5,7-azulene by using the current method. Although there is an even-length path connecting the atoms 1 and 3 (left) and the atoms 5 and 7 (right), the electric current circulates by using any existing odd-length path to avoid the blockage of the current due to lack of polarity alternation (see Figure 2).

In a molecule without odd cycles, i.e., a molecule with bipartite or non-frustrated structure, it is enough to connect the electrodes to a pair of atoms separated by an even number of atoms in order to obtain QI. That is, because there are no odd cycles, all the other paths connecting these two atoms are of even length, such that no alternation of the polarity of the bonds is allowed and the current cannot be transmitted. The prototype of this system is the benzene molecule, which is illustrated in Figure 4.

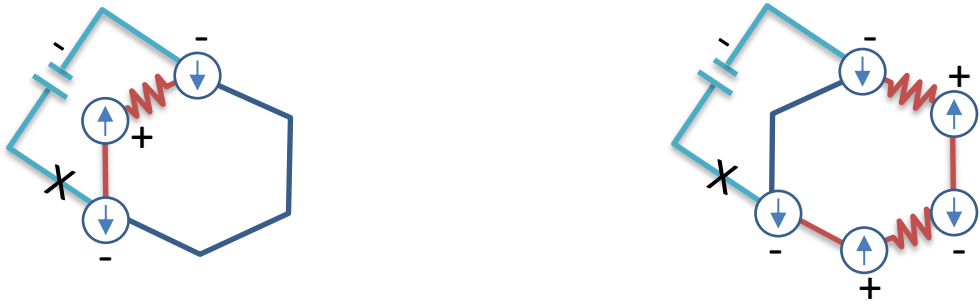

**Figure 4**. Illustration of the QI in 1,3-benzene due to the lack of any odd-length path connecting these two atoms. The current flow is impeded in either of the two existing paths that connect the atoms 1 and 3 due to the lack of polarity alternation in the circuit, such that no transmission is possible between these two atoms.

It has been recently remarked [41] that the sign of $G^0(E_F=0) = -\hat{H}^{-1}$ can play an important role in a yet not reported quantum interference involving more than two atoms. Using the function $\mathscr{F}$ it is now easy to formulate some rules about the sign pattern of $\mathscr{F}_{rs}$. That is,

- $\mathscr{F}_{rs} > 0$ if an only if there is at least one path of length $4k+3$, $k=1,2,\cdots$ between the atoms $r$ and $s$;
- $\mathscr{F}_{rs} < 0$ if an only if there is at least one path of length $4k+1$, $k=1,2,\cdots$ between the atoms $r$ and $s$.

For instance, in benzene the pair 1,2 is connected by a path of length one and by a path of length 5, which are both of the type $4k+1$. Thus, $\mathscr{F}_{1,2} < 0$. On the other hand, the pair 1,4 is connected by two paths of length 3 and $\mathscr{F}_{1,4} > 0$.

Finally, in Table 1 we illustrate the results obtained with $\mathscr{F}_{rs}$ for predicting the conductance in naphthalene and azulene and compare them with the experimental values of the conductance and with those obtained by using $G_{rs}^0(E_F=0)$.

| Naphthalene | | | |
|---|---|---|---|
| Substituent | $G_0$ (exp.)[a,b] | $\left[G_{rs}^0(E_F=0)\right]^2$ | $\left[\mathscr{F}_{rs}(\gamma^2=0.1)\right]^2$ |
| 1,4 | 11.0 | 0.445 | 0.783 |
| 1,5 | 2.2 | 0.111 | 0.274 |
| 2,6 | 1.4 | 0.111 | 0.192 |
| 2,7 | 0.1 | 0 | 0 |
| Azulene | | | |
| 2,6 | 32 | 0.25 | 1.430 |
| 1,3 | 32[c] | 0.25 | 0.934 |
| 4,7 | 8 | 0.25 | 0.720 |
| 5,7 | 2 | 0 | 0.025 |

[a]Taniguchi et al. [27].

[b]Xia et al. [38].

[c]For the pyridine-linked derivative the value of $G_0 = 9$ is reported in [38].

As can be seen $\mathscr{F}_{rs}$ not only predicts correctly the sites where QI appears but also the correct trend of conductance for those sites for which there is a current flow. Notice that although in naphthalene there is a clear trend between conductance and the interatomic distance, which has been previously reported by Taniguchi et al. [27], such trend does not exist for azulene. In fact, for azulene the smallest conductance is observed for the pair 5,7- which is significantly closer to each other than the pair 2,6- which displays a significantly higher conductance. The function $\mathscr{F}_{rs}$ reproduces very well this trend, which is not accounted for by the zeroth Green's function $G_{rs}^0(E_F = 0)$.

# Epilogue

At the very dawn of the XXI century, in 2002, Betowski et al. used post-Hartree-Fock calculations with Configuration Interaction (CI) to study triplet excitation energies of a series of polycyclic aromatic hydrocarbons (PAHs). The study uses a combination of sophisticated *ab initio* techniques with basis sets CIS/6-311G(d,p), CISD/3-21G and UHF-RHF/6-311G(d,p). The paper called my attention because of the calculation time. For perylene, $C_{20}H_{12}$, the calculation took 14 hours when using CIS/6-311G(d,p) on a Cray C94 supercomputer. The time is extended to 24 days if the CISD/3-21G is used. This prompted me to paraphrase the advice that Pliny the Younger gave to his friend Cornelius Rufus [49]: '*I counsel you in that ample and thriving retreat of yours, to hand the repeating and boring calculations over to the computer, and to devote yourself to the study of the mathematical and theoretical aspects of chemistry so as to derive from it something totally your own.*' In [50] we used the simple HMO-GT approach to study the same problem, which at the end of the day was to understand the phototoxicity of these PAHs. The correlation coefficient obtained for the phototoxicity of PAHs based on the HMO-GT is 0.968, which is very much comparable with that obtained using the CIS/6-311G(d,p) basis (0.978). The HMO-GT method gives much better results than the semiempirical methods such as AM1 (0.899) and PM3 (0.904). It is not needed to say that the calculation time using HMO-GT is just a few seconds using any laptop commercially available today.

The reason for telling this anecdote is to recall the necessity of using an Occam Razor in deciding the complexity of the methods to be used for tackling chemical problems. In the dawn of the XXI century it seems to many that employing brute force by using much faster computers will solve all the problems in modern society. The recently coined and fashionable term "Big Data" is the main paradigm of this trend. But sometimes, simple, mathematically elegant and computationally efficient approaches solve the problem in much easier, efficient and elegant way. Thus, in the dawn of the XXI century we should not get rid of the HMO-GT approach for solving increasingly complex theoretical chemistry problems. The method has passed all the challenging tests during its more than 85 years of existence.

## Acknowledgements
The author thanks Professor Roald Hoffmann for introducing him to this topic as well as for useful discussions. Dr Yuta Tsuji is thanked for the preparation of the Figure 2. Professor Michele Benzi is also thanked for useful discussions about the Cayley-Hamilton theorem. The Royal Society of London is thanked for a Wolfson Research Merit Award to the author.